\documentstyle[aps,preprint,epsf]{revtex}

\begin{document}

\draft

\tightenlines

\title{Ground State Properties of Many-Body Systems
in the Two-Body Random Ensemble and Random Matrix Theory}

\author{L. F. Santos \footnote{E-mail:
santos@nst4.physics.yale.edu}, Dimitri Kusnezov \footnote{E-mail:
dimitri@mirage.physics.yale.edu}}

\address{Center for Theoretical Physics,
Sloane Physics Lab, Yale University, New Haven - CT 06520-8120,
USA}

\author{Ph.
Jacquod \footnote{E-mail: pjacquod@lorentz.leidenuniv.nl}}

\address{Instituut-Lorentz, Universiteit Leiden, P.O. Box 9506,
2300 RA Leiden, The Netherlands}

\maketitle

\begin{abstract}
We explore generic ground-state and low-energy statistical properties 
of many-body bosonic and fermionic one- and two-body random ensembles 
(TBRE) in the dense limit, and contrast them with 
Random Matrix Theory (RMT). 
Weak differences in distribution tails can
be attributed to the regularity or chaoticity of the
corresponding Hamiltonians rather than the particle statistics. 
We finally show the universality of
the distribution of the angular momentum gap between the lowest
energy levels in consecutive $J$-sectors for the four models
considered.
\end{abstract}

\pacs{PACS: 05.30.-d, 47.52.+j, 21.60.-n, 21.10.Re, 75.75.+a}

\newpage

Wigner introduced the Gaussian orthogonal ensemble (GOE) to deal
with the statistics of high-lying levels of many-body quantum
systems \cite{Wigner}. Although the fluctuations of certain
observables predicted by the GOE agree well with experimental
observations, it represents a system in which all the particles
interact simultaneously and this is a priori not appropriate to 
describe
many-body systems in which the two-body interaction is
predominant\cite{Weide,Brody}. The two-body random ensemble (TBRE) was 
introduced to improve upon the physical limitations of 
RMT\cite{French}. 
The TBRE Hamiltonian includes only up to two-body operators, 
whose coefficients are real random numbers. This ensemble reproduces a
Gaussian level density and the GOE level repulsion as
desired, but until recently, only the {\sl dilute limit} was 
analytically tractable
\cite{Kota}. This limit corresponds to having a large number of
particles ($N_p \gg 1 $) and an even larger number of
single-particle states $K \gg 1$, so that $N_p/K \ll 1$. In such a
limit Pauli's principle has only a marginal effect, so
particles statistics are unimportant. Many physical regimes lie outside
the dilute limit: for bosonic particles, we may
consider a large number of particles with a finite $K \ll N_p$;
similarily, fermionic antisymmetry becomes relevant as one
approaches half-filling, $N_p \approx K/2$. Since we are interested
in those aspects, we do not restrict ourselves to the dilute limit. It 
has
only recently been shown, from the connection between the Lanczos
tridiagonalization and random polynomials, that this case can also be 
treated
analytically in some instances\cite{Kusnezov}.

We investigate the statistical properties of the low-lying
levels of interacting complex systems, extracting from the edge of
the spectra, generic properties that would reflect
the basic structure of many-body systems such as molecules,
atomic nuclei or quantum dots close to, or in their ground-state.
We compute the lowest eigenvalue distribution for two
bosonic and two fermionic TBREs and show that, after a proper
rescaling, the distributions have a surprisingly
weak dependence on the nature of the
particles. Comparing to predictions from RMT lead to surprising 
similarities.
The fact that the distribution of the lowest energy state
depends only weakly on the nature of the particles and
not on the ensemble used to study it
points to its universality, but a closer inspection of
the tails of these distributions indicate that they depend on the
chaoticity of the system analyzed. To understand the role of chaos, 
we study models which are regular as well as chaotic and find
that the tails of the ground state distributions for
chaotic systems are closer to that obtained from the large-N GOE, 
while for integrable systems it is closer to the N=2 GOE. 
The distribution of energy differences between
the lowest energy levels in two consecutive
angular momentum sector ($J$ and $J+1$) is found to be more robust, 
having the
same shape in all four cases
within numerical accuracy (deviations occur for the single $j$-shell 
model
at small gap values). Since this distribution depends on correlations 
between Hilbert
subspaces with different quantum numbers, it is outside the scope of 
RMT.
We expect that this distribution is generic, 
a conclusion which is borne out by the bosonic
nature of the corresponding excitation.
Our analysis is based on numerical and analytical results. The
analytical treatment is now possible not only for the TBRE 
\cite{Kusnezov},
as already pointed out, but also for the large N limit of the GOE, for 
which an
expression for the distribution function of the largest eigenvalue
was obtained in terms of a particular Painlev\'e II 
function\cite{Tracy}.

We start by summarizing analytical results obtained
for the GOE. A general matrix $H$ of this ensemble can be expressed
in terms of a rotation matrix $O$ and the diagonal eigenvalue
matrix $E = \mbox{diag} (E_{1}, E_{2},... E_{N}) $ as $H = O^{T} E
O $. The distribution of the ground state energy is 
$P(E_{0} ) =\langle\delta (E_{0} -E_{1} )\rangle$, where 
$\langle\cdots\rangle$
indicates GOE averaging and the probability of having a matrix
element $H_{ij}$ is $P(H_{ij} ) \propto \exp (- \frac{1}{2}
\mbox{Tr} H^2 )$. Two limits are considered: $N=2$ and
$N\rightarrow \infty $. For $N=2$, direct integration
gives

\begin{equation}
P_{2}^{GOE}(E_{0} ) = \frac{1}{2\sqrt{\pi }} \left\{ \exp(- E_{0}^2 ) -
\sqrt{\frac{\pi }{2}}  \exp(- E_{0}^2 /2) E_{0}  \left[ \mbox{erfc
} \left( \frac{E_{0} }{\sqrt{2}}\right) \right] \right\} .
\label{N2}
\end{equation}

This distribution is not readily calculated for arbitrary $N$, however 
in the limit 
$N \rightarrow \infty $, Tracy and Widom\cite{Tracy} derived an 
expression for the
distribution of the largest eigenvalue ( $f_1(s)$\cite{Tracy}), 
which in terms of the lowest eigenvalue is written as

\begin{equation}
P_{\infty}^{GOE}(E_{0} ) = \frac{1}{2} 
  \left( \int _{-E_{0}} ^{\infty } q(x)^2 dx
+ q(-E_{0}) \right) \exp \left[ - \frac{1}{2} \left( \int
_{-E_{0}} ^{\infty } (x + E_{0}) q(x)^2 dx + \int _{- E_{0}}
^{\infty } q(x) dx \right)\right] . \label{Tracy}
\end{equation}
Here $q(E_{0})$ satisfies the Painlev\'e II equation and the
boundary condition is $q(E_{0}) \sim \mbox{Ai} (E_{0})$ as $E_{0}
\rightarrow \infty$, where $\mbox{Ai} $ is the Airy function. The
numerical solutions for this expression are plotted on the top panel of
Fig.1, where one sees that it agrees well with the distribution
for a GOE of $N=200$. This function has the asymptotics
\begin{equation}
 \log P_{\infty}^{GOE}(E_{0} ) \sim -\mid E_0\mid^3,
                \quad E_0\rightarrow -\infty,\qquad\qquad
\log P_{\infty}^{GOE}(E_{0} ) \sim -E_0^{3/2},
                \qquad E_0\rightarrow \infty.
\end{equation}
In contrast to $N=2$, the tails deviate from Gaussian behavior.
For completeness, we also show the average ground state energy for the 
GOE,
although this quantity is not expected to have physical consequences.
We find that $\langle E_{0} \rangle = 2.01(2) \sqrt{N} + 1.58(2) 
N^{-1/6}$, 
which agrees with the predicted functional form\cite{Tracy,Sinai}.
On the bottom panel of Fig.1, we show that this expression fits 
numerical results for $N \in [2,1000]$. 

In contrast to the GOE, the TBRE is not a purely statistical model,
but rather is constructed in a particular bosonic or fermionic space.
We take it in the form

\begin{equation}
H = \sum _{ij} \epsilon _{ij} a^{+}_{i} a_{j} + \sum _{ijkl}
V_{ijkl} a^{+}_{i} a^{+}_{j}a_{k} a_l .
\end{equation}
When originally introduced, the Hamiltonian
operators of the TBRE had only (interaction) two-body terms, but we 
include the one-body term.
$a^{+}_{i}$$(a_{i}) $ represents boson or fermion creation
(annihilation) operator and the coefficients
$\epsilon _{ij}$ and $V_{ijkl}$ are taken as Gaussian random
variables once certain physical constraints are imposed such as
rotation invariance, time reversal invariance and conservation of
total spin, isospin and parity.

As already mentioned, some analytical results for bosonic
\cite{Kota,bosondilute} and fermionic \cite{Brody,French} models
in the dilute limit were obtained, but here we consider the
non-dilute limit, for which little is known. We want to
find out how this limit and some other model dependencies, such as
the nature of the particles or the system's integrability, may
affect the statistical properties of many-body systems.

\underline{Bosonic Models}

We study two bosonic models, the U(4) vibron and U(6) interacting
boson models, used to explore molecular or nuclear collective
excitations. In the U(4) TBRE model, bosons with $J^{\pi } =
0^{+}$ ($s$, $s^{+}$) and  $J^{\pi } = 1^{-} $ ($\tilde{p}$,
$p^{+} $) are coupled to form a scalar Hamiltonian \cite{vibron}

\begin{eqnarray}
\label{vibron}
 H_{U(4)} &=& (\epsilon _{s} s^{+} s + \epsilon _{p}
p^{+} . \tilde{p} )/N_p
\\
  &+& \{ \frac{c_{0} }{2} [p^{+} p^{+} ]^{(0)} .
  [\tilde{p} \tilde{p} ]^{(0)} + \frac{c_{2} }{2}
  [p^{+} p^{+}]^{(2)}
  .[\tilde{p} \tilde{p} ]^{(2)}
  +\frac{u_{0} }{2} [s^{+} s^{+} ]^{(0)} .
  [ss]^{(0)}  \nonumber \\
  &+&
  \frac{u_{1} }{2} [s^{+} p^{+} ]^{(1)} . [\tilde{s} \tilde{p}]^{(1)}
  + \frac{v_{0} }{2 \sqrt{2} } ([s^{+} s^{+} ]^{(0)} .
  [\tilde{p} \tilde{p} ]^{(0)} + h.c.)
  \} /N_p(N_p-1) \nonumber .
\end{eqnarray}
The square brackets denote angular momentum coupling, the dots
represent scalar products and $N_p$ is the total number of bosons.
Since the 1- and 2-body matrix elements are proportional to $N_p $
and $N_p (N_p-1)$ respectively, scaling allows all coefficients in
(\ref{vibron}) to be Gaussian random numbers of unit variance.
For the choice of coefficients above,  the lowest eigenvalue 
distribution has been obtained
analytically, and in the large $N_p$ limit\cite{Kusnezov}

\begin{equation}
P(E_{0}) \propto e^{- 0.4 E_{0}^2} \mbox{erfc} (0.663 E_{0}) +
0.423  e^{- E_{0}^2 /2.14 } \mbox{erfc} ( 0.632 E_{0}).
\label{TBRE}
\end{equation}
The second model we investigate is the
nuclear U(6) model (IBM) which consists of scalar $J^{\pi } = 0^{+}$
($s$, $s^{+}$) and quadrupole $J^{\pi } =2^{+}$ ($\tilde{d}$,
$d^{+}$) bosons coupled by one- and two-body interactions.
The Hamiltonian has the form \cite{Iachello}

\begin{eqnarray}
H_{U(6)} &=& (\epsilon _{s} s^{+} s + \epsilon _{d} d^{+} .
\tilde{d} )/N_p
\nonumber \\
  &+& \{ \sum_{L=0,2,4} c_{L} [d^{+} d^{+} ]^{(L)} . [\tilde{d} 
\tilde{d} ]^{(L)}
  + a_{1} [s^{+} d^{+} ]^{(2)} . [\tilde{d} \tilde{d} ]^{(2)} + a_{2}
  [s^{+} s^{+} ]^{(0)} . [ss]^{(0)} \nonumber \\
  &+& a_{3} [s^{+} d^{+} ]^{(2)} . [\tilde{s} \tilde{d} ]^{(2)}
  + a_{4} [s^{+} s^{+} ]^{(0)} . [\tilde{d} \tilde{d} ]^{(0)} + h.c. 
\}/N_p(N_p-1),
\end{eqnarray}
We will consider $N_p=16$, which is typical for a heavy collective
nucleus, and the distribution for the ground state energy is
obtained numerically. An important distinction between the $U(4)$ and 
$U(6)$ models is 
that the former is integrable for all parameters, while the latter is 
generally chaotic\cite{ibmchaos}.
Chaos usually applies to fluctuation properties near the middle of the 
spectrum, rather than at the
edges. Never the less, there is a subtle difference between these 
models which we attribute to the
underlying chaos, as we discuss below. The ground states of these 
models also have well known 
phase transition behaviors. In spite of this, the ground state 
distributions will be seen to be 
surprisingly insensitive to this.

\underline{Fermionic Models}

The two fermionic models we investigate are a model for
interacting nucleons in a single j-shell and a model for
randomly interacting electrons. For the single j-shell we use
$N_p=6$ particles in the $j=15/2$ shell. The matrix elements of
the Hamiltonian are \cite{Talmi}

\begin{eqnarray}
& & <j^{N_p} \alpha J M |\sum _{i<k}^{N_p} V_{ik} | j^{N_p} \alpha
' J M
> \nonumber \\
&=& \frac{N_p(N_p-1)}{2} \sum _{\alpha " ,J", J'} [j^{N_p} \alpha
J \{ | j^{N_p-2} (\alpha " J") j^{2} (J') J][j^{N_p-2} (\alpha "
J") j^{2} (J') J| \} j^{N_p} \alpha ' J] v_{J'},
\end{eqnarray}
where $J$ is the total angular momentum, $M$ is the $z$-projection,
$\alpha $ denotes all other quantum numbers, the term in
brackets corresponds to the coefficients of fractional parentage
and $v_{J'}$ are random numbers. 

The second fermionic model is a model for randomly
interacting $j=1/2$ fermions (IEM) and is given by \cite{IEM}

\begin{equation}
H = \sum_{\alpha j} \epsilon _{\alpha } c^{\dag }_{\alpha
j} c_{\alpha j} + \sum _{\alpha \beta \gamma \eta j j' }
U^{\gamma \eta }_{\alpha \beta } c^{\dag }_{\alpha j}
c^{\dag }_{\beta j'}c_{\gamma j'} c_{\eta j},
\end{equation}
where the one-body spectrum is Wigner-Dyson distributed with
$\epsilon _{\alpha } \in [-K/2;K/2]$ ($K/2$ gives the number
of spin-degenerate orbitals, the average level spacing is
then $\Delta = 1$), $U^{\gamma \eta }_{\alpha \beta } \in [-U;U]$
are Gaussian random numbers and $j (j') = \uparrow (\downarrow)$
are spin indices. We study up to $N_p=7$ fermions on $K/2=7$
orbitals.

While the single j-shell model describes non-chaotic system \cite{Zel}, 
the
spectral properties of the IEM depend on the ratio of the
interaction strength to the one-body level spacing $U/\Delta$: for
$U/\Delta \ll 1/(K N_{p}^{2})$ the model is non-chaotic with a
Poissonian spectrum while for $U/\Delta \gg 1/(K N_{p}^{2})$ it is
chaotic with a Wigner-Dyson distributed spectrum \cite{Dima}.

To compare the lowest eigenvalue distribution for these four
models with the ones for the GOE, we re-center the distributions
and rescale their widths by using the following scaling
prescription

\begin{equation}
\varepsilon_0 = \frac{E_{0} - <E_{0}>}{\sigma _{E_{0}}}.
\end{equation}
This removes model dependencies associated with scales, level densities 
and so forth.
We show in Fig.2 distributions of the rescaled ground state energy
$\varepsilon_0$ for the four models described above. The figure
splits into regular systems on the top two panels
and chaotic ones on the two bottom panels.
Bosonic models are left and fermionic ones are right. 
The chaoticity of the IEM depends on
$U/\Delta$ and the data shown on the lower right panel of Fig.2
for the IEM correspond to $U/\Delta=1$, well into the chaotic
regime \cite{Dima}.

One clearly sees that the ground state energies have approximately
the same distribution for the four TBREs and the GOE of small and
large dimension - this seems to be a robust property, and in
particular, these distributions seem not to depend on the nature
of the particles. Moreover, the agreement of the distributions
with the ones obtained from the GOE demonstrates that the distribution
of the ground state energy is also not sensitive to the type of
ensemble used. A closer inspection of the presented data does reveal
discrepancies between different models in the distribution tails.
The lowest energy distribution for regular systems coincides
better with the distribution for the two-dimensional GOE, while
the distribution for chaotic systems is in better agreement with the 
one
obtained with a large-N GOE. The GOE dimension necessary
to describe a system can be associated with the integrability of the
model. This conclusion is corroborated by the data shown on
Fig.3 for the IEM at various $U/\Delta$, both in the integrable and 
chaotic
regimes. Evidently the crossover from Poisson to Wigner-Dyson
distributed level spacing (i.e. from integrable to chaotic
quantum dynamics) is accompanied by a crossover from
two-dimensional to infinite-dimensional GOE fitting of the
ground-state distribution.
We also found that the ground-state distribution for the IEM
exactly at half-filling ($K=14$ and $N_p=7$), gets close to a
two-dimensional GOE, as shown at the top of Fig.4. Since the
associated spectrum is still Wigner-Dyson distributed, we conclude
that this is a manifestation of the particle statistics. The only
influence of the latter therefore is to single out half-filled
fermionic systems which despite their chaotic character have a
distribution of ground-state energies which is closer to the $N=2$
GOE.

Consider now distributions which cannot be addressed through RMT, 
namely
distributions which involve correlations between Hilbert subspaces with 
different
quantum numbers. One quantity of fundamental interest is the energy 
gap. 
In the models we study, this
typically involves states of different spin:   for the vibron model
it is $\varepsilon(J^{\pi} = 1^{-}) - \varepsilon(J^{\pi } = 0^{+} )$, 
for
the IBM  $\varepsilon(J^{\pi } = 2^{+}) - \varepsilon(J^{\pi } =
0^{+} )$, while for the single $j$-shell model and the IEM
$\varepsilon(J=1)-\varepsilon(J=0)$. We will call this energy
difference the {\it spin gap}. The ground state is mostly
dominated by $0^{+}$ (this predominance almost reaches 100\% for
the IEM \cite{IEM}) and the lowest level with a different spin
usually corresponding to the first excited state (this is however
not the case for the IEM), so we are looking at the very edge of
the spectrum. Even though we have both bosonic and fermionic
models, the distributions are very similar and also agree very
well with the analytical expression derived  for the
vibron model \cite{Kusnezov} as it is evident from Fig.5.
The distribution of the spin gaps is thus also a
generic property which we attribute to the bosonic nature (i.e.
$\delta J$ is an integer) of the corresponding excitation in all
four models studied here. This explains why the particle
statistics have no influence here. Note that in the case of the
IEM, universality of the spin gap requires a finite interaction
strength $U/\Delta \gtrsim 1$. In the limit $U/\Delta \rightarrow
0$, the spin gap is determined by the one-body spectrum
$\epsilon_{\alpha}$ and thus given by a GOE one- (two-) spacing
distribution for $N_p$ even (odd).

In summary, investigating various TBREs, we have found that in
spite of their fundamental differences, they have strikingly
similar features in both their ground state energy distribution
and their spin gap distribution. The ground state energy
distributions have common features, and coincide to  GOE distributions
of either small or large dimensions, for regular or chaotic TBREs
respectively. We conclude that the edge of the spectra exhibit two
unexpected generic properties: the distribution of the ground
state energy and the distribution of the energy differences
between the lowest levels with different spins. This latter point
is presumably related to the apparent regular structure found in
the angular momentum structure of low-lying levels for several
TBREs \cite{gsspin}.

\acknowledgments L. F. Santos acknowledges the support of the
Funda\c c\~ao de Amparo \`a Pesquisa do Estado de S\~ao Paulo
(FAPESP) and Ph. Jacquod has been supported by the Swiss National
Science Foundation.

\begin{figure}
\epsfxsize=5.5in
\epsffile{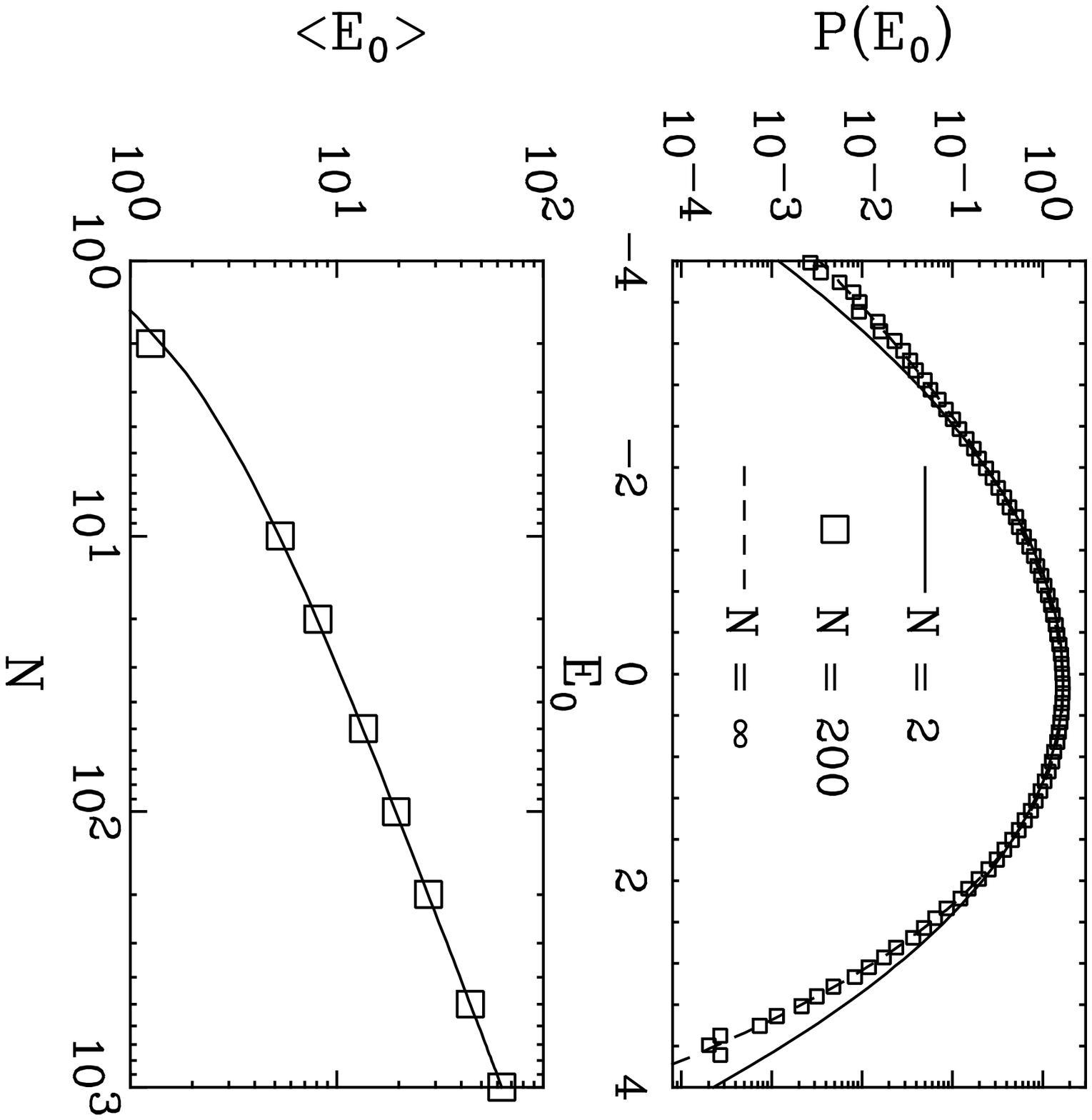}

\caption{Top: Analytical distribution of the ground state energy
for GOE of dimension $N=2$ (solid line) and $N=\infty$ (dashed line).
The numerically obtained distribution for $N=200$ (squares)
already coincides with the $N=\infty$ GOE. Bottom: Average
GOE ground-state energy $\langle E_0 \rangle$ as a function of
the matrix size $N$. The solid line gives the theoretical
estimate.} \label{fig1}
\end{figure}

\begin{figure}[ht]
\epsfxsize=6in
\epsffile{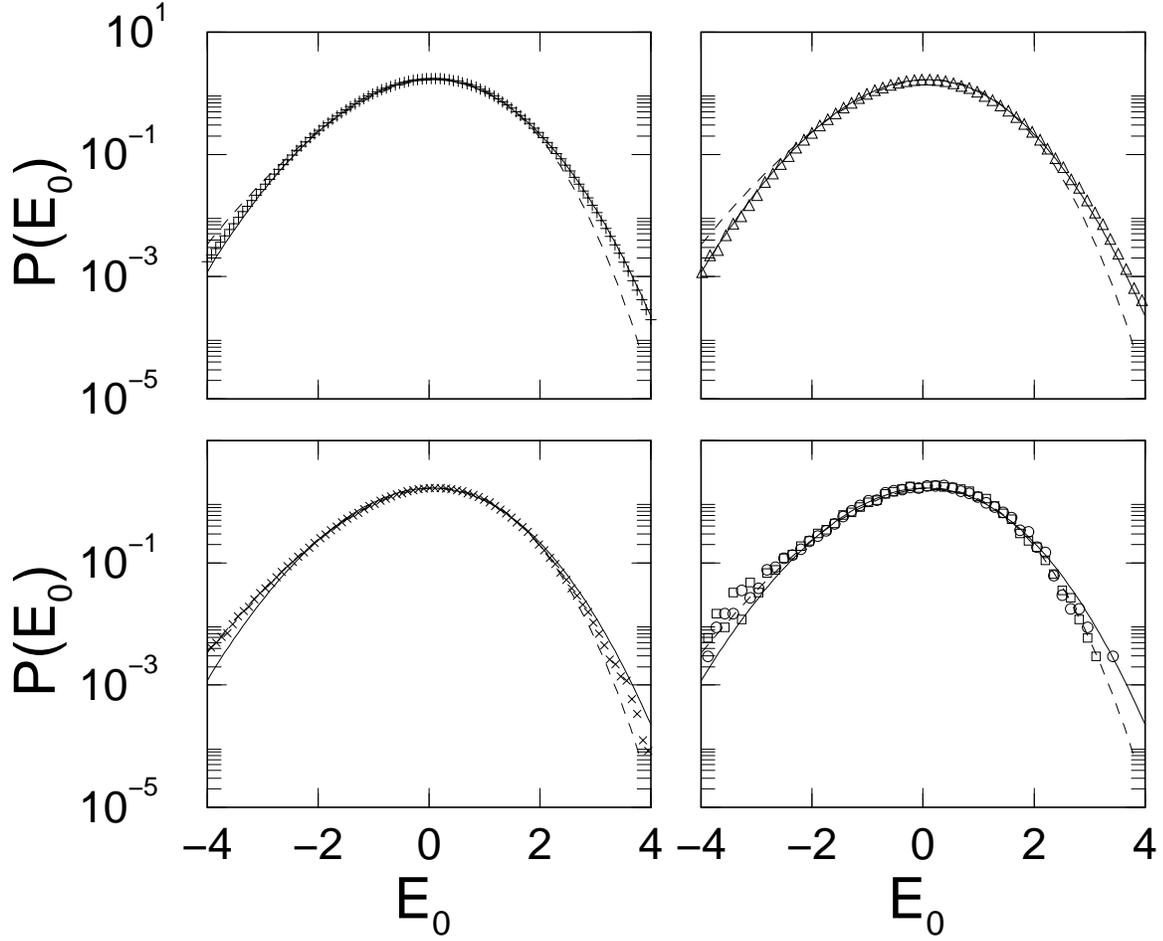}
\vspace{1cm}
\caption{Distributions of the ground state energy for the vibron
model (top left), the single $j$-shell model (top right), the IBM
(bottom left) and the IEM (bottom right) for $N_p=3$ (circles) and
4 (squares). In all panels, solid and dashed lines give the
distribution for the $N=2$ and $N=\infty$ GOE respectively. The
top panels correspond to non-chaotic systems, for which the data
are well fitted by a $N=2$ GOE. The bottom panels correspond to
chaotic systems, where data are closer to the $N=\infty$ GOE.}
\label{fig2}
\end{figure}

\begin{figure}[ht]
\epsfxsize=4.5in
\epsffile{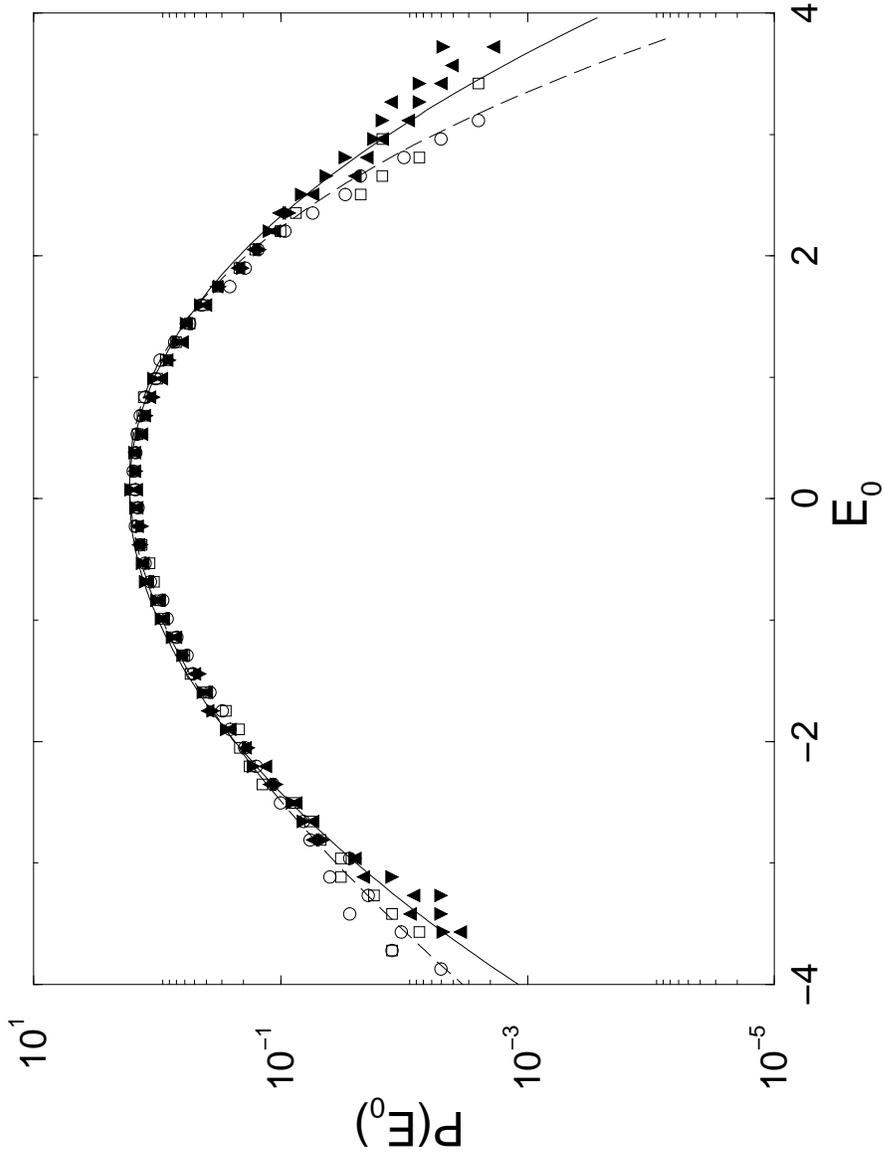}
\vspace{1.5cm}
\caption{Crossover to quantum chaoticity for the IEM with $N_p=4$
and $U/\Delta=10^{-7}$ (diamonds), $0.05$ (triangles), $1$
(squares) and $\infty$ (circles). Symbols for data in the
integrable regime are filled, while solid and dashed lines give
the distribution for the $N=2$ and $N=\infty$ GOE
respectively.}\label{fig3}
\end{figure}

\begin{figure}[ht]
\epsfxsize=4.5in
\epsffile{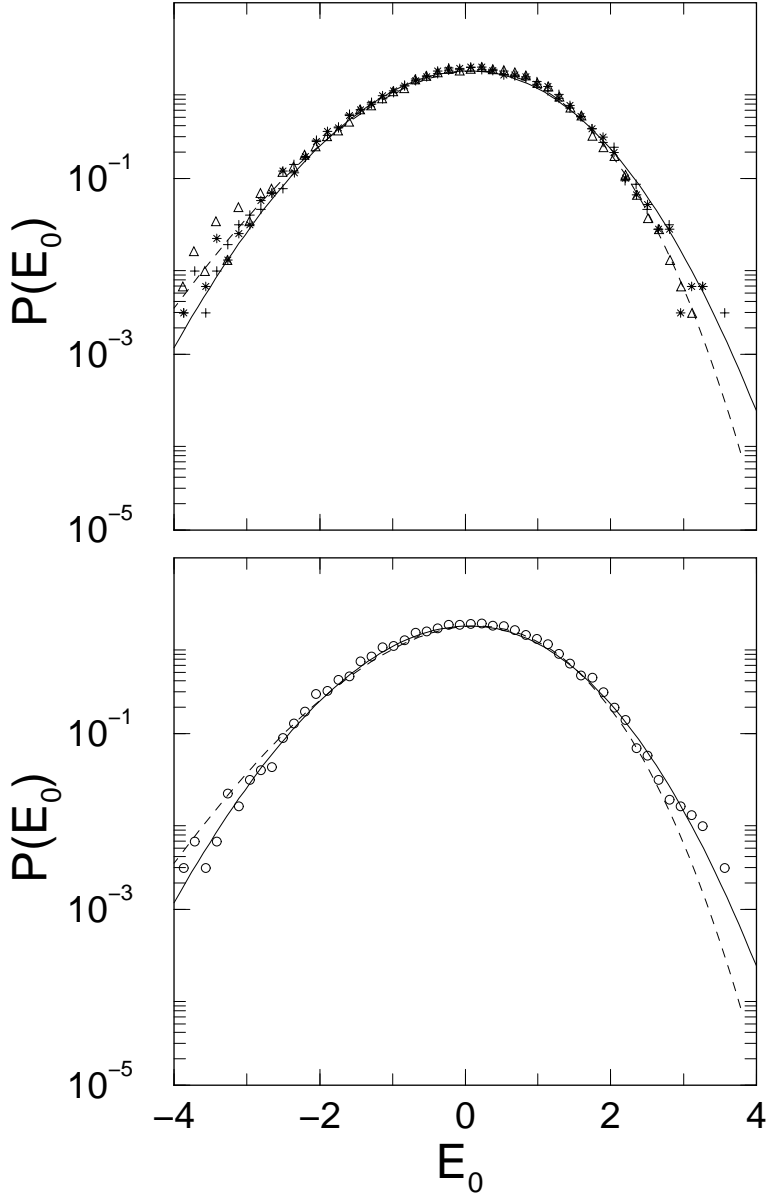}
\vspace{1cm}
\caption{Top panel: Distribution of the ground state energy for
the IEM for $N_p=4$ (triangles), 5 (crosses) and 6 (stars). Bottom
panel: Distribution of the ground state energy for the IEM at half
filling, $N_p=7$ (circles). In both panels, solid and dashed lines
give the distribution for the $N=2$ and $N=\infty$ GOE
respectively. The particle statistics plays a role at half-filling
only, rendering the distribution closer to the $N=2$ GOE.}
\label{fig4}
\end{figure}

\begin{figure}[ht]
\epsfxsize=5.5in \epsffile{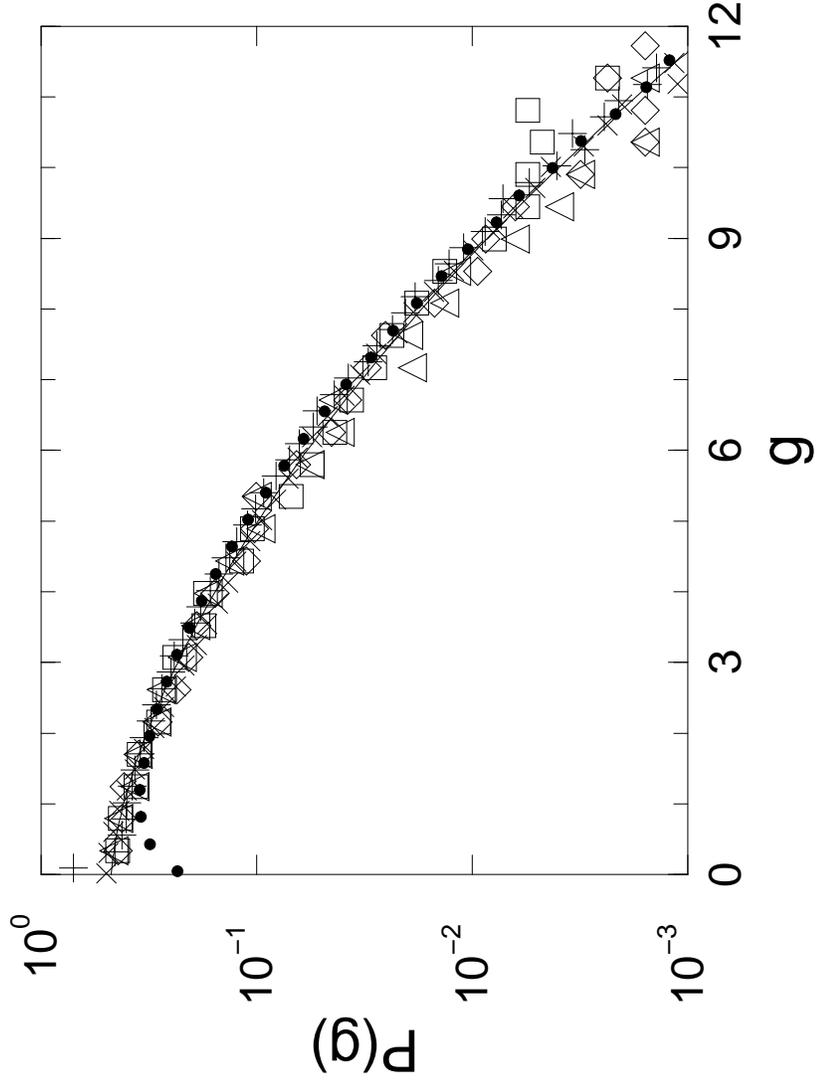}
\caption{Spin gap
distribution for the U(4) vibron model ($\times$) the U(6) IBM (+)
the single $j$-shell model ($\bullet$) and the IEM with $N_P=5$
(triangles), 6 (squares) and 7 (diamonds). The solid line
corresponds to the analytical expression obtained for the vibron
model. The only significant deviations occur for the single
$j$-shell model for low gap values.} \label{fig5}
\end{figure}

\end{document}